\documentclass[sigconf]{acmart}
\usepackage{subcaption}
\usepackage{listings}
\usepackage{xcolor}
\usepackage{multirow}
\usepackage{enumitem}
\usepackage[normalem]{ulem}
\usepackage{makecell}
\newcommand{\OMIT}[1]{}
\lstdefinelanguage{SQL}{
  morekeywords={
   SELECT, FROM, WHERE, GRAPH_TABLE, MATCH, SOURCE, KEY, DESTINATION, REFERENCES, LABEL, NODE, EDGE, VERTEX, TABLES, CREAT, PROPERTY, GRAPH, RETURN, JOIN, PROPERTIES, ON, DISTINCT, WITH, AS, UNION, ALL, EXISTS, ANY, RECURSIVE, ORDER, BY, AND
  },
  sensitive=false,
  morecomment=[l]{--},
  morestring=[b]',
}
\newcommand{\sqlkw}[1]{\texttt{\textcolor{blue}{#1}}}

\usepackage{graphicx}
\usepackage{booktabs}
\renewcommand\footnotetextcopyrightpermission[1]{}  

\newcommand{\mcomment}[2]{{\color{blue}\textbf{(#1)}}\footnote{\textbf{#1:} #2}}
\newcommand{\liat}[1]{\mcomment{Liat}{#1}}

\copyrightyear{2025}
\acmYear{2025}
\setcopyright{cc}
\setcctype{by}
\acmConference[GRADES-NDA '25]{8th Joint Workshop on Graph Data Management Experiences \& Systems (GRADES) and Network Data Analytics (NDA) }{June 22--27, 2025}{Berlin, Germany}
\acmBooktitle{8th Joint Workshop on Graph Data Management Experiences \& Systems (GRADES) and Network Data Analytics (NDA) (GRADES-NDA '25), June 22--27, 2025, Berlin, Germany}\acmDOI{10.1145/3735546.3735860}
\acmISBN{979-8-4007-1923-3/2025/06}

\title{Towards Cross-Model Efficiency in SQL/PGQ}

\author{Hadar Rotschield}
\affiliation{%
 \institution{School of CS, Hebrew University}
 \city{Jerusalem}
 \country{Israel}
}
\email{hadar.rotschield@mail.huji.ac.il}

\author{Liat Peterfreund}
\affiliation{%
 \institution{School of CS, Hebrew University}
 \city{Jerusalem}
 \country{Israel}
}
\email{liat.peterfreund@mail.huji.ac.il}

\begin{document}

\begin{abstract}
SQL/PGQ is a new standard that integrates graph querying into relational systems, allowing users to freely switch between graph patterns and SQL. 
Our experiments show performance gaps between these models, as queries written in both formalisms can exhibit varying performance depending on the formalism used, suggesting that current approaches handle each query type separately, applying distinct optimizations to each formalism.
We argue that a holistic optimization is necessary, where the system internally decides on the best algorithms regardless of whether queries are written in SQL or as graph patterns. We propose possible future research direction to unify these optimizations and mitigate performance gaps.

\end{abstract}



\maketitle

\section{Introduction}
Modern database systems increasingly support both relational and graph queries, as reflected in the emerging SQL/PGQ standard. SQL/PGQ is part of the broader GQL initiative~\cite{GQL}, which has been under development since 2019 by both academia and industry contributors under the ISO auspices. GQL consists of two parts: a standalone graph query language, namely GQL, and SQL/PGQ, which extends SQL with constructs for querying property graphs.

SQL/PGQ allows to define property graph views over relational data and query them using a pattern-matching syntax inspired by ASCII art. These graph queries return relational results, which can then be further processed using standard SQL~\cite{icdt23,pods23}. 
This gives the flexibility to move between graph and relational modeling depending on the query. For example, reachability is often easier to express using graph traversal, while aggregation is more natural in SQL.

Despite this flexibility, it is unclear how the choice of model affects performance. Ideally, the user's choice, whether to use SQL or graph patterns, should be driven by clarity and convenience, not performance concerns. For example, pattern-matching queries that could be rewritten as joins should benefit from existing join optimization techniques~\cite{wcoj,ngo2014skew}. Similarly, recursive SQL queries might perform better when viewed and executed as graph traversals. In other words, the way a query is written should not significantly affect its efficiency. 
Since query equivalence testing is generally undecidable, the goal should be feasable translations, regardless of using SQL or SQL/PGQ.
In this work, we present  a short set of experiments that compare the performance of SQL and SQL/PGQ queries across several systems. Our study focuses on three questions: (1) Does the query language impact performance? (2) Can relational optimizations be reused in the graph setting? (3) Should graph views be fully materialized or kept virtual?

We ran a set of experiments on several query types using DuckDB with the DuckPGQ extension~\cite{DuckPGQ2023}, and ran comparative tests on Google Cloud Spanner~\cite{GoogleCloudSpanner} and Neo4j~\cite{Neo4j_docs}. Our results show that performance is often tied to how the query is expressed, suggesting that the two models are not yet fully decoupled. We argue that achieving such decoupling, where the system chooses the best execution plan regardless of query syntax, is key to making hybrid querying efficient.

To move toward this goal, we outline two directions for future research: (1) applying and combining proven algorithmic approaches from both relational and graph processing to obtain performance guarantees, and (2) enabling internal query rewriting, so the system can automatically translate queries between models when beneficial. 
\OMIT{
\begin{figure}[b]
\centering
\begin{lstlisting}[language=SQL, numbers=none, caption = Graph View Creation, label = q:graph, captionpos=b]
CREATE PROPERTY GRAPH social_graph
    VERTEX TABLES (
        Person
          PROPERTIES (pid, name, city)
          LABEL "Person",
        Account
          PROPERTIES (aid, type)
          LABEL "Account"
    )
    EDGE TABLES (
        Friend
          SOURCE KEY (pid1) REFERENCES Person (pid)
          DESTINATION KEY (pid2) REFERENCES Person (pid)
          PROPERTIES (since)
          LABEL "Friend",
        Owns
          SOURCE KEY (pid) REFERENCES Person (pid)
          DESTINATION KEY (aid) REFERENCES Account (aid)
          LABEL "Owns",
        Transfer
          SOURCE KEY (from) REFERENCES Account (aid)
          DESTINATION KEY (to) REFERENCES Account (aid)
          PROPERTIES (amount)
          LABEL "Transfer"
    );
\end{lstlisting}
\end{figure}
}
\paragraph{Related Work}
Classical work on query optimization has focused on finding efficient join orders~\cite{Ibaraki1984,Krishnamurthy1986,Kossmann2022}. 
Graph pattern matching has been extensively studied~\cite{Angles2017}. In sequential settings, Ullmann’s backtracking algorithm~\cite{Ullmann1976} has been optimized using trie indexing~\cite{Shang2008}, symmetry breaking~\cite{Han2013}, and compression~\cite{Bi2016}. 

Cross-model query optimization is gaining popularity in both research and industry. Recent work has expanded traditional Select-Project-Join pipelines to include graph operators for more efficient neighbor lookups~\cite{Lou2024}. 
\OMIT{There is also growing interest in combining relational and graph models: DuckPGQ~\cite{Wolde2023CIDR, DuckPGQ2023} embeds SQL/PGQ in DuckDB~\cite{DuckDB2024} to process pattern matching. Index-based solutions such as GQ-Fast~\cite{Lin2016} and GRainDB~\cite{Jin2022} build graph-like indexes over relational data to speed up joins. ReIGo uses GRainDB’s indexing for graph operations, and approaches such as GRFusion~\cite{Hassan2018} and Gart~\cite{Shen2023} materialize graph data from relational sources, allowing direct graph query execution, but at the cost of more storage and potential data inconsistencies.
}

\section{Querying Property Graphs in SQL/PGQ}
\label{sec:Querying_Property_Graphs_in_SQL/PGQ}

SQL/PGQ extends SQL to support property graph querying by defining graphs as views over relational data. It uses pattern matching to find patterns in those graphs, producing relations as output. This output can be further processed with SQL.

\subsection*{Property Graphs as Relational Views}
Property graphs store data in a flexible graph format, allowing labels and key-value properties on both nodes and edges. PGQ enables to construct such graphs from relational data.
Let us consider the relational schema that consists of the following relations with underlined primary keys.

\begin{itemize}
  \item \texttt{Person}(\underline{pid}, name, city)
  
  \item \texttt{Account}(\underline{aid}, type)

  \item \texttt{Own}(pid, aid)
  \begin{itemize}
    \item \textit{pid} is a foreign key → \texttt{Person}(pid)
    \item \textit{aid} is a foreign key → \texttt{Account}(id)
  \end{itemize}

  \item \texttt{Friends}(pid1, pid2, since)
  \begin{itemize}
    \item \textit{pid1}, \textit{pid2} are foreign keys → \texttt{Person}(pid)
  \end{itemize}

  \item \texttt{Transfer}(\underline{tid}, from, to, amount)
  \begin{itemize}
    \item \textit{from}, \textit{to} are foreign keys → \texttt{Account}(aid)
  \end{itemize}
\end{itemize}

The following SQL/PGQ code constructs a property graph with Person and Account as nodes, and Friends (between people), Owns (person to account), and Transfer (between accounts) as edges.
\begin{lstlisting}[language=SQL, numbers=none]
CREATE PROPERTY GRAPH social_graph
    VERTEX TABLES (
        Person
          PROPERTIES (pid, name, city)
          LABEL "Person",
        Account
          PROPERTIES (aid, type)
          LABEL "Account"  )
    EDGE TABLES (
        Friend
          SOURCE KEY (pid1) REFERENCES Person (pid)
          DESTINATION KEY (pid2) REFERENCES Person (pid)
          PROPERTIES (since)
          LABEL "Friend",
        Owns
          SOURCE KEY (pid) REFERENCES Person (pid)
          DESTINATION KEY (aid) REFERENCES Account (aid)
          LABEL "Owns",
        Transfer
          SOURCE KEY (from) REFERENCES Account (aid)
          DESTINATION KEY (to) REFERENCES Account (aid)
          PROPERTIES (amount)
          LABEL "Transfer" );
\end{lstlisting}

\OMIT{
\begin{figure*}[t!]
    \centering
    \includegraphics[width=0.8\textwidth]{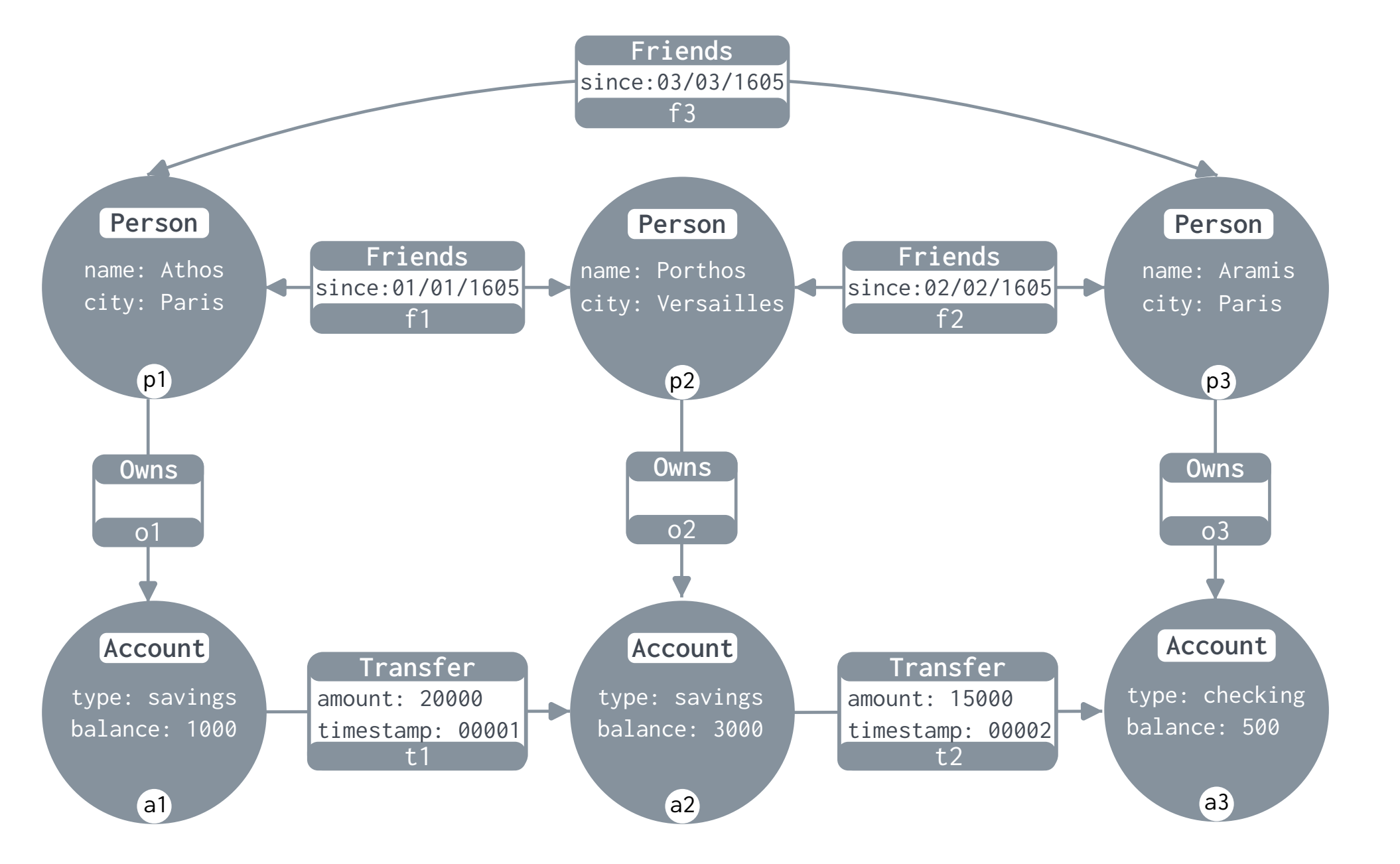}
    \caption{An example of a labeled property graph taken from \cite{TheoreticalModels}.}
    \label{fig:property-graph-example}
\end{figure*}
}

\subsection*{Pattern Matching}
On such property graph views, we can apply pattern matching which extracts relations from graphs. \emph{Graph patterns}, formalized in~\cite{icdt23,pods23}, are specified using ASCII-art-like syntax to describe structures in the graph. 
 We distinguish between two types of patterns - bounded and unbounded. We refer to a pattern as \emph{bounded} when it  has a bounded length (e.g., a triangle of exactly three edges). 
Such queries can often be rewritten to SQL using standard joins, without recursion. 
In contrast, an \emph{unbounded} pattern uses Kleene's star, hence allowing arbitrarily long edge traversal (e.g., a path of any length), 
typically requiring recursion in SQL.
Here, bounded Query~\ref{query1} identifies triangle of friends. 
\begin{lstlisting}[language=SQL, numbers=none, caption = Bounded Friends Triangle, label = query1, captionpos=b]
SELECT *
FROM GRAPH_TABLE (
  social_graph
  MATCH 
    (x:"Person") -[f1:Friend]-> (y:"Person"),
    (y:"Person") -[f2:Friend]-> (z:"Person"),
    (z:"Person") -[f3:Friend]-> (x:"Person")
  RETURN (x.name, y.name, z.name); );
\end{lstlisting}

Each line after the~\sqlkw{MATCH} specifies a directed friendship edge between two Person nodes. The commas between the lines act like joins. (Full syntax and semantics are in~\cite{pods23}.)
We can also add the following lines before~\sqlkw{RETURN} to verify that $x$ transferred money to $y$ and add filtering to detect suspicious money transfers: 
\begin{lstlisting}[language=SQL, numbers=none]
    (x:"Person") -[:Owns]-> (ax:"Account"),
    (y:"Person") -[:Owns]-> (ay:"Account"),
    (z:"Person") -[:Owns]-> (az:"Account"),
    (ax:"Account") -[t1:Transfer]-> (ay:"Account"),
    (ay:"Account") -[t2:Transfer]-> (az:"Account"),
    (az:"Account") -[t3:Transfer]-> (ax:"Account")
  WHERE
    x.city = y.city AND x.city != z.city 
                    AND t1.amount > t2.amount
\end{lstlisting}
In Query~\ref{query4}, we detect unbounded cycles of bank transfers with Kleene's star. That is, the `\lstinline|->*|' matches an unbounded sequence of edges. In particular, $x$ transfer to $y$, $y$ to $z$, and there is an unbounded sequence of transfers starting from $z$ and ending in $x$.
\begin{figure}[ht!]
\centering
\begin{lstlisting}[language=SQL, numbers=none, caption = Unbounded Transfers Cycle, label = query4, captionpos=b]
SELECT *
FROM GRAPH_TABLE (
  social_graph
  MATCH p = ANY SHORTEST
    (x:"Account") -[t1:Transfer]-> (z:"Account")
                  -[t2:Transfer]-> (y:"Account")
                  -[t3:Transfer]-> *(x:"Account")
    RETURN; );
\end{lstlisting}
\Description{}
\end{figure}

Notice that we use \texttt{ANY SHORTEST} to restrict to a shortest cycle that satisfies the pattern. The reason behind this is to avoid infinite loops when trying to detect the cycle. 
We can also add a filter:
\begin{lstlisting}[language=SQL, numbers=none]
  WHERE px.city <> pz.city AND t1.amount > t2.amount
\end{lstlisting}

This query introduces two filters before the \sqlkw{RETURN}. 
\begin{itemize}
    \item \texttt{px.city <> pz.city} ensures that the owners of \texttt{x} and \texttt{z} are from different cities.
    \item \texttt{t1.amount > t2.amount} restricts the path to cases where the first transfer amount is greater than the second.
\end{itemize}
Such constraints are useful for detecting specific patterns, such as suspicious circular transfers, by combining structural and property-based conditions.
Note that queries using the Kleene star can be rewritten as recursive SQL. However, the reverse is not true: there are queries expressible with recursive SQL that cannot be expressed in SQL/PGQ~\cite{TheoreticalModels}.

\begin{table*}[t]
\centering
\begin{tabular}{|c|cc|cc|cc|c|c|c|}
\hline
\textbf{Dataset} & \multicolumn{2}{c|}{\textbf{Q1}} & \multicolumn{2}{c|}{\textbf{Q2}} & \multicolumn{2}{c|}{\textbf{Q3}} & \textbf{Q4} & \textbf{Q5} & \textbf{Q6} \\
\textbf{Size (\# rows)} & \textbf{DuckDB} & \textbf{Spanner} & \textbf{DuckDB} & \textbf{Spanner} & \textbf{DuckDB} & \textbf{Spanner} & \textbf{DuckDB} & \textbf{DuckDB} & \textbf{DuckDB} \\
\hline
50  & 1.11 & 0.50  & 1.72 & 0.16  & 1.28 & 0.04  & 41.67 & 2.53   & 1.92  \\
100 & 0.33 & 0.33  & 2.16 & 0.14  & 1.96 & 0.16  & 0.32  & 0.84   & 1.95  \\
150 & 1.61 & 0.49  & 0.95 & 0.16  & 1.30 & 0.045 & 7.75  & 250.00 & 1.03  \\
\hline
\end{tabular}
\caption{SQL Execution Time Divided by SQL/PGQ Time: DuckDB vs. Spanner}
\label{tab:merged-factor-diffs}
\end{table*}
\section{From SQL/PGQ to (Recursive) SQL}
SQL/PGQ queries that do not use Kleene's closure (i.e., no unbounded edge traversals) can be rewritten as basic SQL without recursion. For instance, Query~\ref{query1} can be expressed in SQL as follows:
\begin{lstlisting}[language=SQL, numbers=none]
WITH FriendPairs AS (
    SELECT pid1 AS person, pid2 AS friend
    FROM Friend
    WHERE pid1 != pid2 
    UNION ALL
    SELECT pid2 AS person, pid1 AS friend
    FROM Friend
    WHERE pid1 != pid2 )
SELECT DISTINCT
    f.pid1,
    f.pid2
FROM Friend AS f
WHERE EXISTS (
    SELECT 1
    FROM FriendPairs fp1
    JOIN FriendPairs fp2 ON fp1.friend = fp2.friend 
    WHERE fp1.person = f.pid1 AND fp2.person = f.pid2 );
\end{lstlisting}

The results in~\cite{figueira2024relationalperspectivegraphquery} indicate that core SQL/PGQ, like core GQL, can be translated into first-order logic with transitive closure. When unbounded repetition is excluded, only basic first-order logic is needed, aligning with standard SQL without recursion. As discussed in Section~\ref{sec:conclusions_and_future_directions}, further work is required to translate these theoretical insights into practical tools and implementations.

When unbounded repetition is involved we need to use recursive SQL. Here, we limit the recursion depth to ensure we do not expand paths indefinitely, mirroring the termination condition of \texttt{ANY SHORTEST}. In practice, this bound guarantees that only short cycles are explored, approximating the behavior of \texttt{ANY SHORTEST} without risking infinite recursion.
\begin{lstlisting}[language=SQL, numbers=none, label = qrec, caption = Recusive Unbounded Friends Cycle,  captionpos=b]
WITH RECURSIVE paths(a_start, a_current, depth) AS (
    SELECT a_from, a_to, 1
    FROM Transfer
    UNION ALL
    SELECT p.a_start, t.a_to, p.depth + 1
    FROM paths p
    JOIN Transfer t ON p.a_current = t.a_from
    -- Limit recursion depth
    WHERE p.depth < 2000 )
SELECT DISTINCT p.a_start AS account_in_cycle
FROM paths p
JOIN Transfer t ON p.a_current = t.a_from
WHERE t.a_to = p.a_start
  AND p.depth >= 2
ORDER BY account_in_cycle;
\end{lstlisting}

Query~\ref{query4} can be expressed with the recursive SQL Query~\ref{qrec}.

\OMIT{
This CTE \texttt{paths} repeatedly expands the search by joining the current node (\texttt{p.a\_current}) to its outgoing edges (\texttt{t.a\_to}). When it finds a row where the destination node equals the start node \texttt{(t.a\_to = p.a\_start)}, it identifies a cycle.
}
Translating from Recursive SQL back to SQL/PGQ is generally not possible due to higher expressiveness of the former~\cite{TheoreticalModels}. Although a complete systematic translation does not exist, certain fragments can still be translated  (see more in Section~\ref{sec:conclusions_and_future_directions}).

\section{Empirical Analysis}
\subsection*{Experimental Setting}
{
We ran six queries (\textbf{Q1}--\textbf{Q6})~\cite{CMESQLPGQRepo} covering bounded friend‑triangles patterns (\textbf{Q1}--\textbf{Q3}) and unbounded multi‑hop transfer patterns (\textbf{Q4}--\textbf{Q6}) on datasets of different sizes. 
We compared SQL and SQL/PGQ latency across DuckDB~\cite{DuckPGQ2023}, and Google Cloud Spanner~\cite{GoogleCloudSpanner}. We checked latency for the same queries in Neo4j's Cypher~\cite{openCypher}. While query syntax varied slightly between systems, all implementations were based on SQL/PGQ.

\paragraph{Datasets Generation}
The datasets of different sizes, 50, 100, and 150 transfers (rows or edges, depends on the model), were generated using \texttt{Mockaroo}~\cite{Mockaroo}. Mockaroo samples each numeric value independently from a uniform distribution over the specified range.
}

\paragraph{Query Scenarios}
We designed six queries for our experiments. The first three involve bounded patterns: they identify pairs of friends connected through a common friend, with increasing levels of complexity by adding financial transactions and city-based constraints. The next three involve unbounded patterns: they detect circular money transfers of arbitrary lengths, some with additional conditions like participants living in different cities or the transfer amounts decreasing.

\paragraph{SQL/PGQ Support Across Systems} We tested SQL/PGQ queries in three systems, DuckPGQ, Google Cloud Spanner, and Neo4j. 
DuckPGQ is an actively evolving extension of DuckDB 
that provides native SQL/PGQ support. 
It translates pattern-matching syntax into relational plans and utilizes 
in-memory data structures for graph operations.
Google Cloud Spanner supports SQL/PGQ through an explicit graph schema layered on top of relational tables, enabling scalable graph querying within its distributed environment. 
We refer to Spanner’s property-graph functionality as “SQL/PGQ” because it adheres to two ISO standards~\cite{GoogleCloudSpanner}:
\begin{itemize}[leftmargin=1em]
    \item ISO/IEC~9075-16:2023 --- \emph{Information technology --- Database languages SQL Property Graph Queries (SQL/PGQ), Edition 1, 2023}
    \item ISO/IEC~39075:2024 --- \emph{Information technology --- Database languages --- GQL, Edition 1, 2024}
\end{itemize}
Neo4j, a native graph database, uses Cypher and optimized data structures for fast traversal, but does not support standard SQL, so only SQL/PGQ results were evaluated.

We conducted our experiments on Neo4j and Google Spanner via their respective web
interfaces, meaning our queries were executed in a cloud-based environment whose
exact hardware configuration was not directly under our control.

For DuckPGQ, we ran all tests locally on a Windows 10 workstation with the following
specifications:
\texttt{Intel i5-8265U} @ 1.60GHz CPUs, 16\,GB RAM.
DuckPGQ was installed using the latest release available at the time.

{These heterogeneous execution settings introduce bias, as cloud latency and resource
provisioning can fluctuate independently of the query engine.  
Thus, we interpret cross system results qualitatively 
rather than drawing conclusions from absolute runtimes.}

\subsection*{Experimental Results}

\label{sec:Results}

\begin{figure}[t]
\centering

\begin{subfigure}{0.3\linewidth}
    \includegraphics[width=\linewidth]{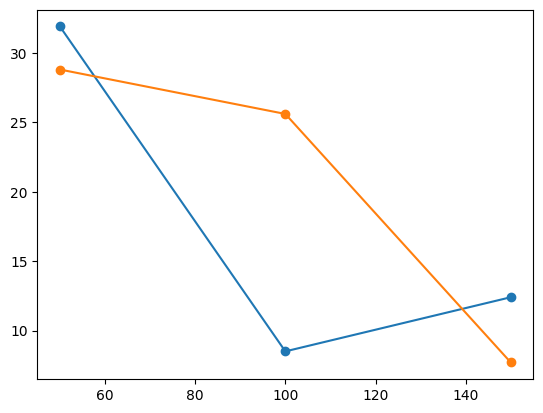}
    \caption{DuckDB: Q1}
    \label{fig:duckdb_bounded_query1}
\end{subfigure}
\hspace{0.3em}
\begin{subfigure}{0.3\linewidth}
    \includegraphics[width=\linewidth]{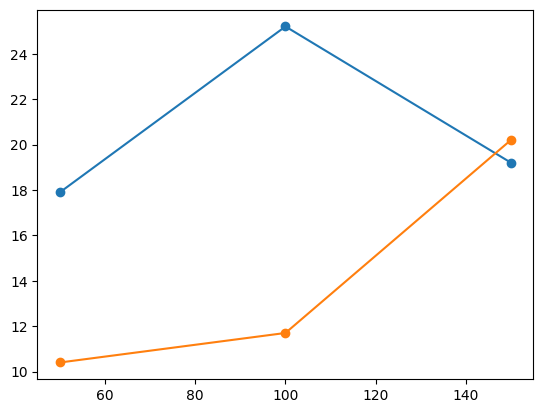}
    \caption{DuckDB: Q2}
    \label{fig:duckdb_bounded_query2}
\end{subfigure}
\hspace{0.3em}
\begin{subfigure}{0.3\linewidth}
    \includegraphics[width=\linewidth]{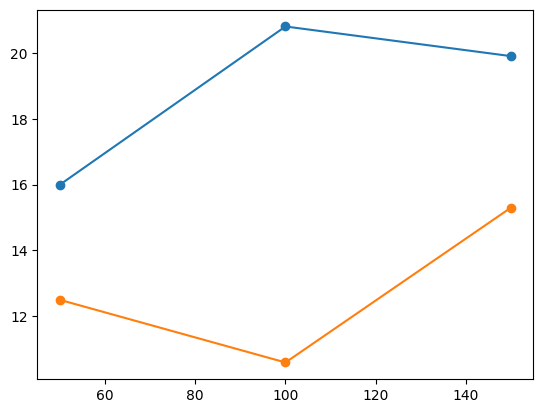}
    \caption{DuckDB: Q3}
    \label{fig:duckdb_bounded_query3}
\end{subfigure}

\begin{subfigure}{0.3\linewidth}
    \includegraphics[width=\linewidth]{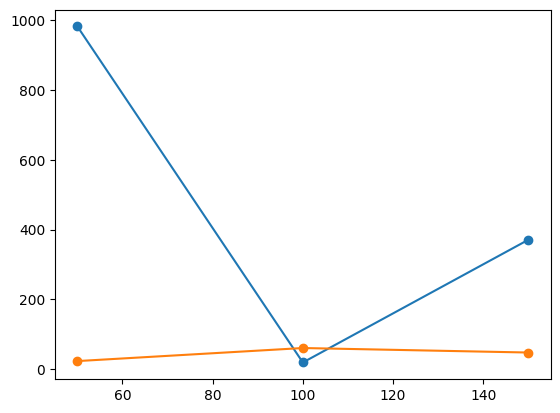}
    \caption{DuckDB: Q4}
    \label{fig:duckdb_unbounded_query1}
\end{subfigure}
\hspace{0.3em}
\begin{subfigure}{0.3\linewidth}
    \includegraphics[width=\linewidth]{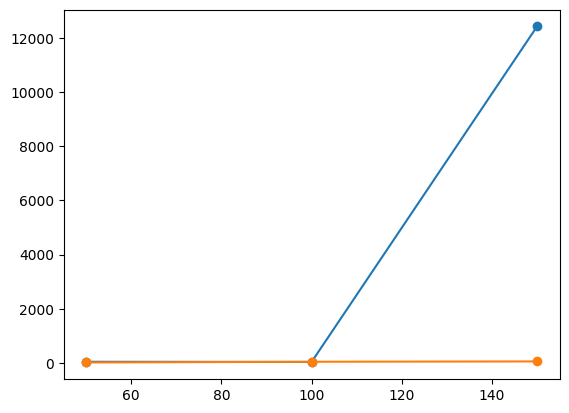}
    \caption{DuckDB: Q5}
    \label{fig:duckdb_unbounded_query2}
\end{subfigure}
\hspace{0.3em}
\begin{subfigure}{0.3\linewidth}
    \includegraphics[width=\linewidth]{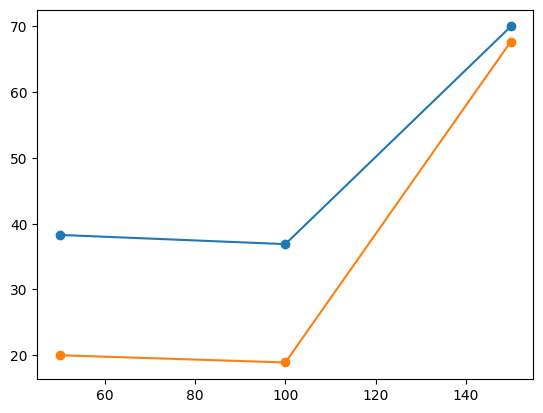}
    \caption{DuckDB: Q6}
    \label{fig:duckdb_unbounded_query3}
\end{subfigure}

\begin{subfigure}{0.3\linewidth}
    \includegraphics[width=\linewidth]{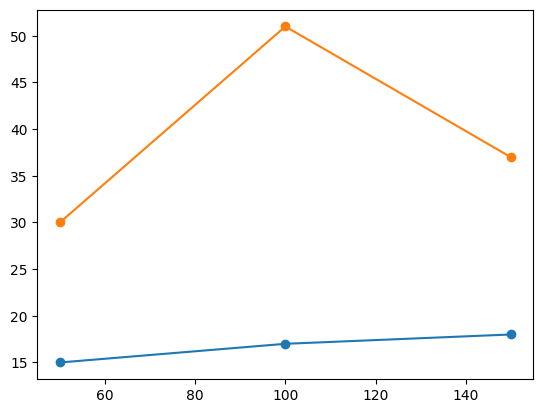}
    \caption{Spanner: Q1}
    \label{fig:spanner_bounded_query1}
\end{subfigure}
\hspace{0.3em}
\begin{subfigure}{0.3\linewidth}
    \includegraphics[width=\linewidth]{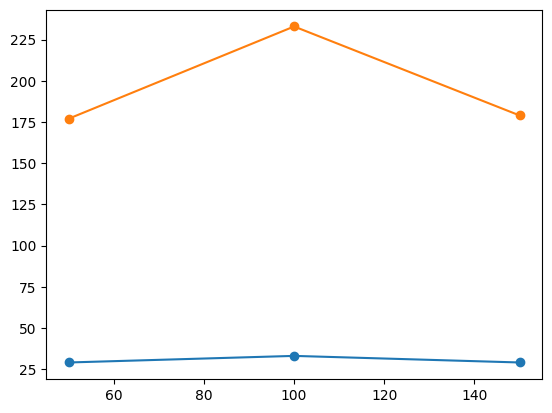}
    \caption{Spanner: Q2}
    \label{fig:spanner_bounded_query2}
\end{subfigure}
\hspace{0.3em}
\begin{subfigure}{0.3\linewidth}
    \includegraphics[width=\linewidth]{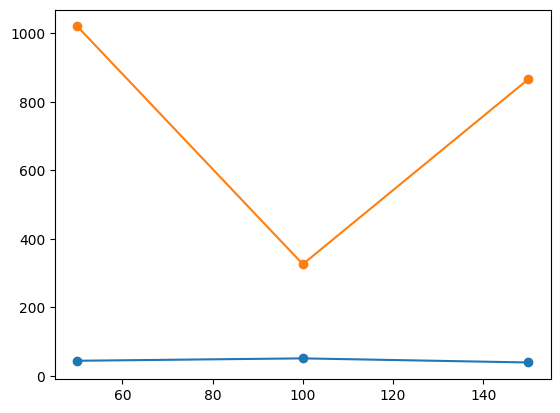}
    \caption{Spanner: Q3}
    \label{fig:spanner_bounded_query3}
\end{subfigure}
\begin{subfigure}{0.15\linewidth}
    \includegraphics[width=\linewidth]{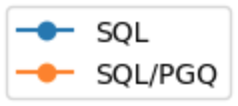}
\end{subfigure}
\caption{Execution time (ms) as a function of dataset size (\#rows) for DuckDB (bounded \& unbounded queries) and Spanner (bounded queries).}
\label{fig:sql_pgq_results}
\Description{}
\end{figure}

\paragraph{SQL vs.~SQL/PGQ in Spanner and DuckDB (Figure~\ref{fig:sql_pgq_results} and Table~\ref{tab:merged-factor-diffs})}

Across bounded queries in Google Cloud Spanner, SQL queries exhibit lower latency than SQL/PGQ (Figures~\ref{fig:sql_pgq_results}(g)-(i)), suggesting stronger optimization for traditional joins. Spanner does not currently support unbounded queries in SQL/PGQ, so we only report bounded queries.
In contrast, DuckDB consistently shows improved or comparable performance for SQL/PGQ, particularly on queries involving unbounded traversals (Figures~\ref{fig:sql_pgq_results}(a)-(f)).

\begin{figure}[t]
\centering
\begin{subfigure}{0.3\linewidth}
    \includegraphics[width=\linewidth]{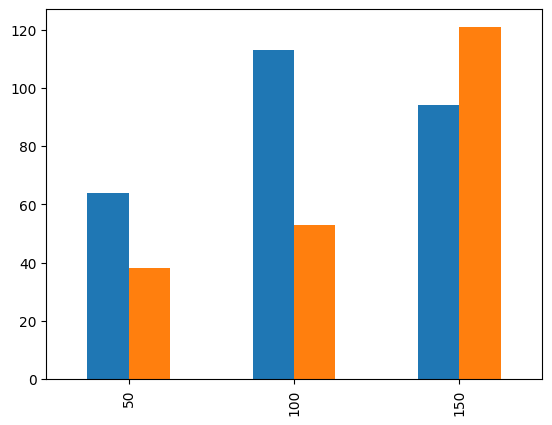}
    \caption{Q1 vs. Q4}
    \label{fig:neo4j_query1}
\end{subfigure}
\hspace{0.3em}
\begin{subfigure}{0.3\linewidth}
    \includegraphics[width=\linewidth]{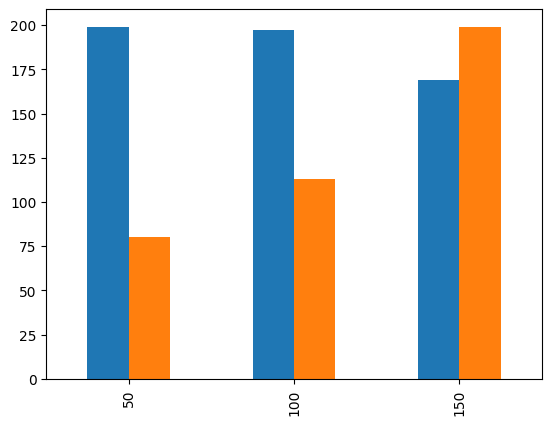}
    \caption{Q2 vs. Q5}
    \label{fig:neo4j_query2}
\end{subfigure}
\hspace{0.3em}
\begin{subfigure}{0.3\linewidth}
    \includegraphics[width=\linewidth]{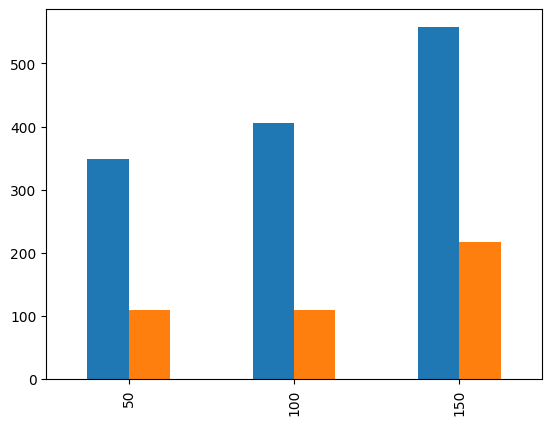}
    \caption{Q3 vs. Q6}
    \label{fig:neo4j_query3}
\end{subfigure}
\begin{subfigure}{0.15\linewidth}
    \includegraphics[width=\linewidth]{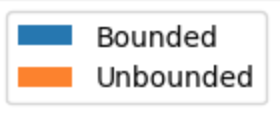}
\end{subfigure}
\caption{Comparison of execution time (ms) as a function of dataset size (\#rows) for Neo4j queries (in Cypher).}
\label{fig:neo4j_results}
\Description{}
\end{figure}

In addition to comparing exact runtimes, which are highly diverse, in Table~\ref{tab:merged-factor-diffs}, we measure the ratio of SQL runtime to SQL/PGQ runtime to provide a more comparative assessment. Specifically,
a value greater than 1 indicates SQL/PGQ is faster.
From the table we see that for DuckDB most factor differences exceed 1, revealing that SQL/PGQ is generally faster than SQL for these queries. In contrast,  for Google Spanner SQL outperforms SQL/PGQ.
These findings highlight that the efficiency of SQL compared to  SQL/PGQ depends on internal optimizations. 
\paragraph{Neo4j (Figure~\ref{fig:neo4j_results})}
Since Neo4j is a native graph database that does not support SQL, we compared bounded vs.~unbounded SQL/PGQ-based queries, where each pair \((\text{Q}i,\text{Q}(i+3))\) is identical except the latter replaces a fixed‑length pattern with a Kleene‐star traversal.  

\paragraph{Graph Creation Times (Table~\ref{tab:graph_creation_latency})}
In DuckDB, constructing graph views on top of relational tables is almost instantaneous. 
Conversely, Neo4j and Spanner report higher overhead, surpassing average query execution latencies by a significant margin. 
This overhead becomes noteworthy in workflows where graph views must be repeatedly initialized.

\OMIT{We see that:\liat{too many details}
\begin{itemize}
    \item \textbf{Query~1} exhibits moderate growth in runtime as dataset size increases, with bounded and unbounded variants sometimes crossing over in performance (e.g., bounded appears faster at certain sizes, slower at others).
    \item \textbf{Query~2} follows a similar pattern, with overall times slightly higher than Query~1.
    \item \textbf{Query~3} shows a larger spread, suggesting that additional graph traversal or filtering (e.g., city constraints, transaction amount checks) in the pattern-matching can notably affect Neo4j's performance.
\end{itemize}
}

\paragraph{Summary} Our results show that DuckDB often benefits from SQL/PGQ compared to plain SQL, regardless of whether queries are bounded
or unbounded. By contrast, Google Cloud Spanner’s performance on bounded SQL queries is generally superior to its SQL/PGQ equivalents.
Neo4j shows mixed performance between bounded and unbounded queries. 
\OMIT{
Although we plotted the raw execution times for each query on each platform, the scales differ significantly across queries and systems. As a result, directly comparing absolute runtimes in a single figure can be misleading. To mitigate this, Tables~\ref{tab:duckdb-factor-diffs} and \ref{tab:spanner-factor-diffs} show the factor differences between the \texttt{SQL} and \texttt{SQL/PGQ} approaches, illustrating more clearly how much faster or slower each approach is relative to the other. No purely SQL baseline is available in Neo4j, so factorized differences are not shown.}


\begin{table}[ht]
\centering
\begin{tabular}{|l|l|l|}
\hline
\makecell{\textbf{Dataset Size} \\ \textbf{(\# rows)}} & \textbf{Platform} & \makecell{\textbf{Graph Creation} \\ \textbf{Latency (ms)}} \\ \hline
\multirow{3}{*}{50}
  & DuckDB         & 0.2   \\
  & Google Spanner & 4,700 \\
  & Neo4j          & 4,614 \\ \hline
\multirow{3}{*}{100}
  & DuckDB         & 0.4   \\
  & Google Spanner & 4,780 \\
  & Neo4j          & 4,668 \\ \hline
\multirow{3}{*}{150}
  & DuckDB         & 0.5   \\
  & Google Spanner & 4,870 \\
  & Neo4j          & 4,293 \\ \hline
\end{tabular}
\caption{Graph Creation Latency by Dataset Size and Platform}
\label{tab:graph_creation_latency}
\end{table}

\section{Conclusion and Future Directions}\label{sec:conclusions_and_future_directions}

The performance differences observed across systems highlight that optimization is primarily determined by how each system optimizes and executes queries internally, rather than by the choice between SQL and SQL/PGQ. Rather than relying on users to select the most efficient query language, systems should aim to automatically optimize queries, regardless of their form. Moving forward, we aim to explore techniques that allow systems to seamlessly translate and optimize queries across models, achieving consistent performance improvements irrespective of query representation.

\paragraph{Theoretical Approaches to Algorithm Optimization}
Join algorithms have been studied and optimized
for many years~\cite{wcoj}, and we can leverage many of those ideas for graph edge traversal. When the traversal is bounded (for example, fixed-length paths), simple join techniques are enough, and there is no need for recursion. For unbounded traversals, indexing strategies like those used in~\cite{wolde2023duckpgq} can help improve performance by avoiding full graph scans. 
On the other hand, we can use graph indexing techniques to develop better algorithms for recursive queries. 

\paragraph{Internal Rewriting}
To reduce performance gaps, systems should automatically rewrite queries between pattern matching and SQL when needed. 
Since SQL is Turing-complete, any SQL/PGQ pattern can, in principle, be translated into SQL. However, the other direction is limited: SQL/PGQ cannot express certain linear recursive queries, as shown in~\cite{TheoreticalModels}. Such translations can occur not only at the high-level query language but also at the intermediate level, such as relational algebra or within the query plan. Query equivalence in relational settings is generally undecidable, and this likely extends to cross-model translations. Therefore, the goal is to identify specific cases where rewritings are feasible and apply them effectively.

\begin{acks}
The authors were supported by ISF grant 2355/24.
\end{acks}

\bibliographystyle{ACM-Reference-Format}
\bibliography{bibliography}

\clearpage      
\onecolumn      
\appendix      

\section{Appendix}
\subsection{Property Graph definitions}

We use the standard definition of property graphs from \cite{TheoreticalModels} where we consider only directed edges. Formally, a property graph is a tuple \(G = \langle N, E, \text{lab}, \text{src}, \text{tgt}, \text{prop} \rangle\), where:
\begin{itemize}
    \item \(N \subseteq \mathbb{N}\) is a finite set of node identifiers used in \(G\).
    \item \(E\subseteq \mathbb{E}\) is a finite set of directed edge identifiers used in \(G\).
    \item \(\text{lab} : N \cup E \rightarrow 2^{\mathcal{L}}\) is a labeling function that associates each node or edge with a finite (possibly empty) set of labels from a set \(\mathcal{L}\).
    \item \(\text{src}, \text{tgt}: E \rightarrow N\) are functions defining the source and target nodes for each edge.
    \item \(\text{prop} : (N \cup E) \times \mathcal{K} \rightarrow \text{Const}\) is a partial function associating nodes or edges with constants (property values), given keys from a set \(\mathcal{K}\).
\end{itemize}
A \emph{path} in a property graph \(G\) is defined as an alternating sequence of nodes and edges \(u_0 e_1 u_1 e_2 \dots e_n u_n\) for \(n \geq 0\), starting and ending with nodes. Each edge \(e_i\) connects nodes \(u_{i-1}\) and \(u_i\). Paths may involve forward edges (from \(u_{i-1}\) to \(u_i\)) or backward edges (from \(u_i\) to \(u_{i-1}\)).
In~\ref{tab:schema-overview} we present the scheme of the relational database with which we experiment with:
\begin{table}[ht]
\centering
\begin{tabular}{p{0.15\linewidth} p{0.7\linewidth}}
\toprule
\textbf{Table} & \textbf{Columns and Description}\\
\midrule
\texttt{Person} 
  & \textbf{Columns:} 
    \textit{pid} (PK), 
    \textit{name}, 
    \textit{city}. \\
\midrule
\texttt{Account} 
  & \textbf{Columns:} 
    \textit{id} (PK), 
    \textit{type}.\\
\midrule
\texttt{Own} 
  & \textbf{Columns:} 
    \textit{pid} (FK $\rightarrow$ \texttt{Person}(pid)), 
    \textit{a\_id} (FK $\rightarrow$ \texttt{Account}(id)). \\
\midrule
\texttt{Friends} 
  & \textbf{Columns:} 
    \textit{pid1} (FK $\rightarrow$ \texttt{Person}(pid)), 
    \textit{pid2} (FK $\rightarrow$ \texttt{Person}(pid)), 
    \textit{since}. \\
\midrule
\texttt{Transfer} 
  & \textbf{Columns:} 
    \textit{tid} (PK), 
    \textit{a\_from} (FK $\rightarrow$ \texttt{Account}(id)), 
    \textit{a\_to} (FK $\rightarrow$ \texttt{Account}(id)), 
    \textit{amount}. \\
\bottomrule
\end{tabular}
\caption{Example schema overview.}
\label{tab:schema-overview}
\end{table}

\begin{figure*}[ht!]
    \centering
    \begin{subfigure}{0.13\textwidth}
        \centering
        \includegraphics[width=\linewidth]{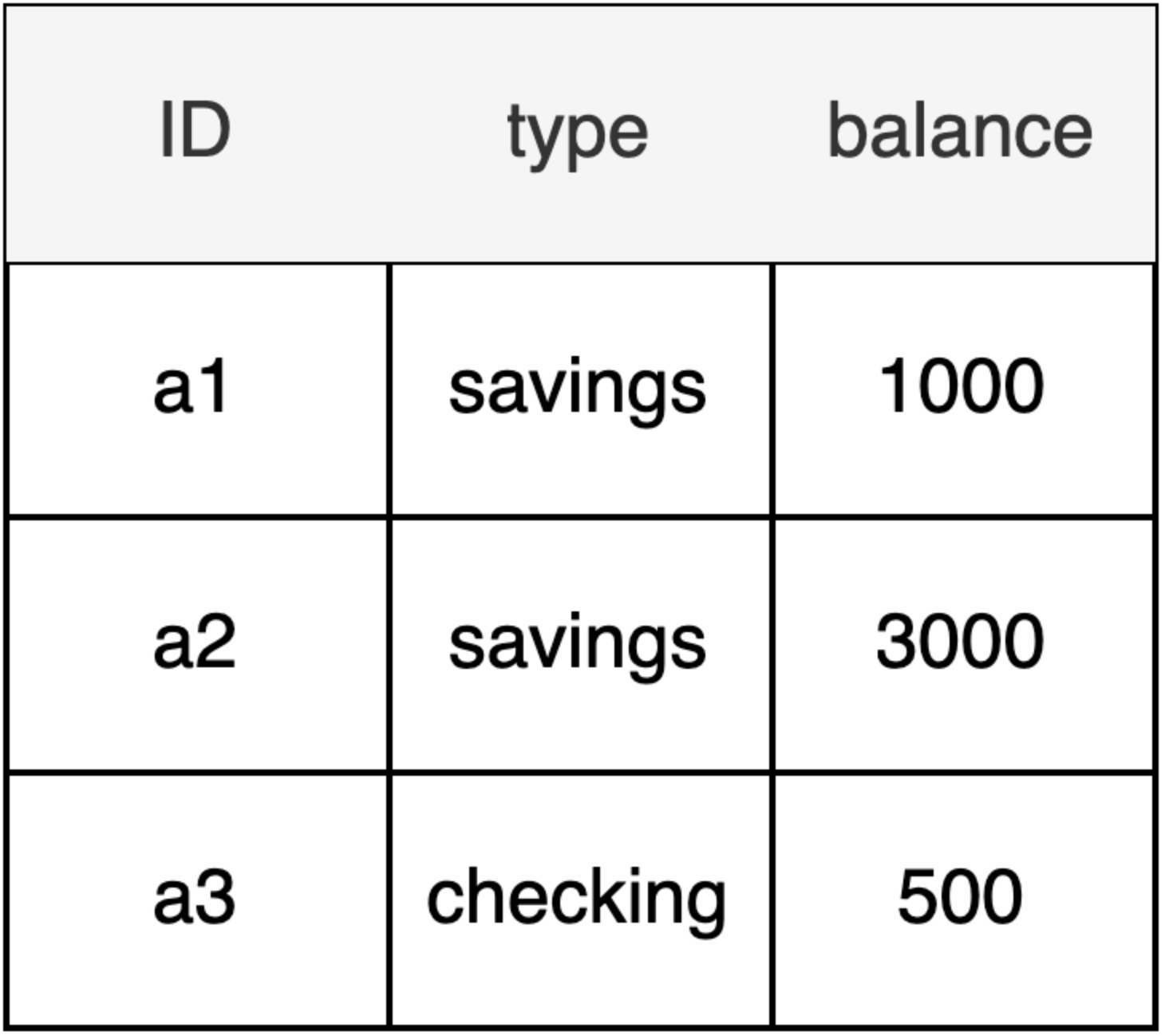}
        \caption{Account}
        \label{fig:account}
    \end{subfigure}
    \hfill
    \begin{subfigure}{0.13\textwidth}
        \centering
        \includegraphics[width=\linewidth]{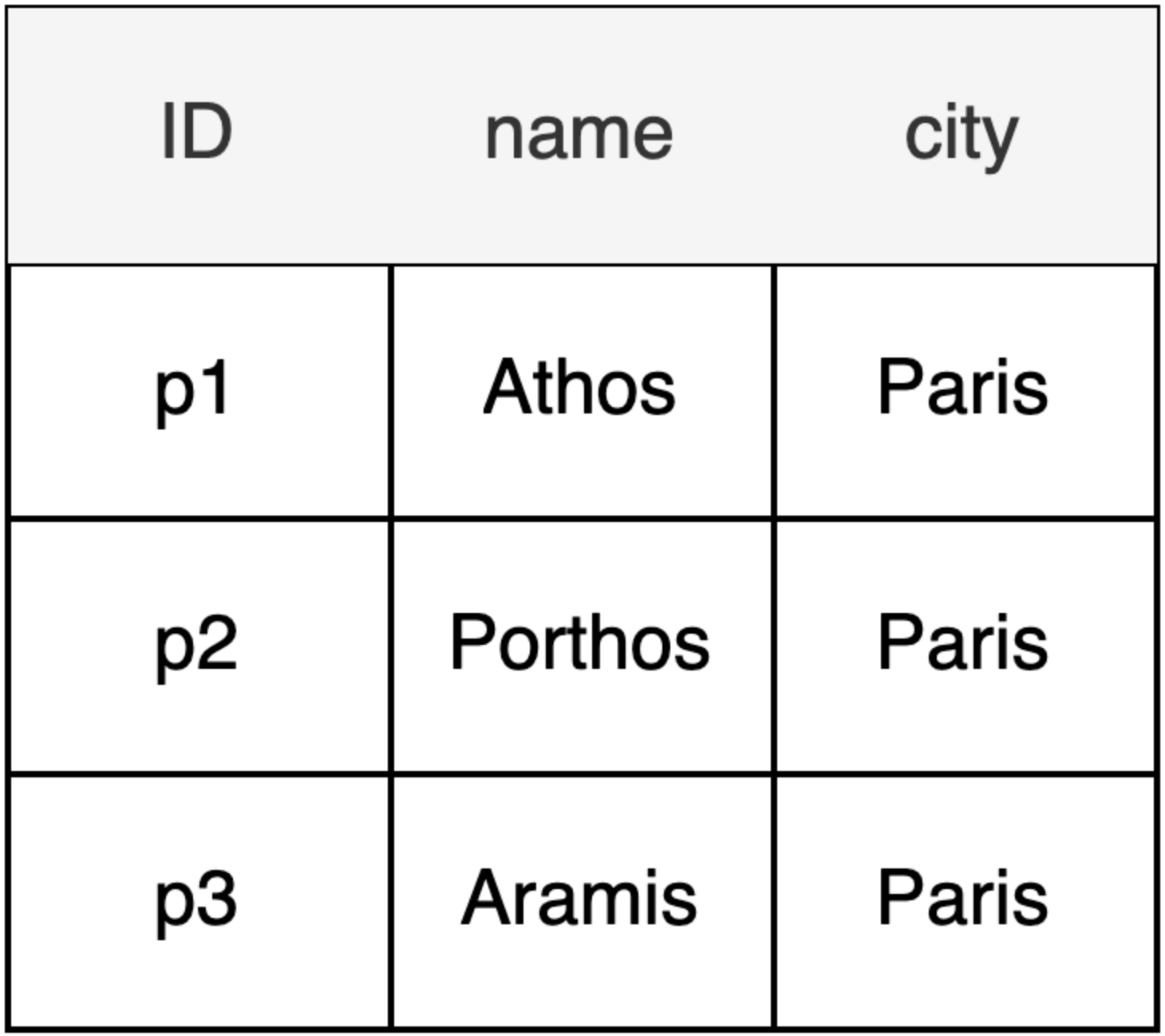}
        \caption{Person}
        \label{fig:person}
    \end{subfigure}
    \hfill
    \begin{subfigure}{0.18\textwidth}
        \centering
        \includegraphics[width=\linewidth]{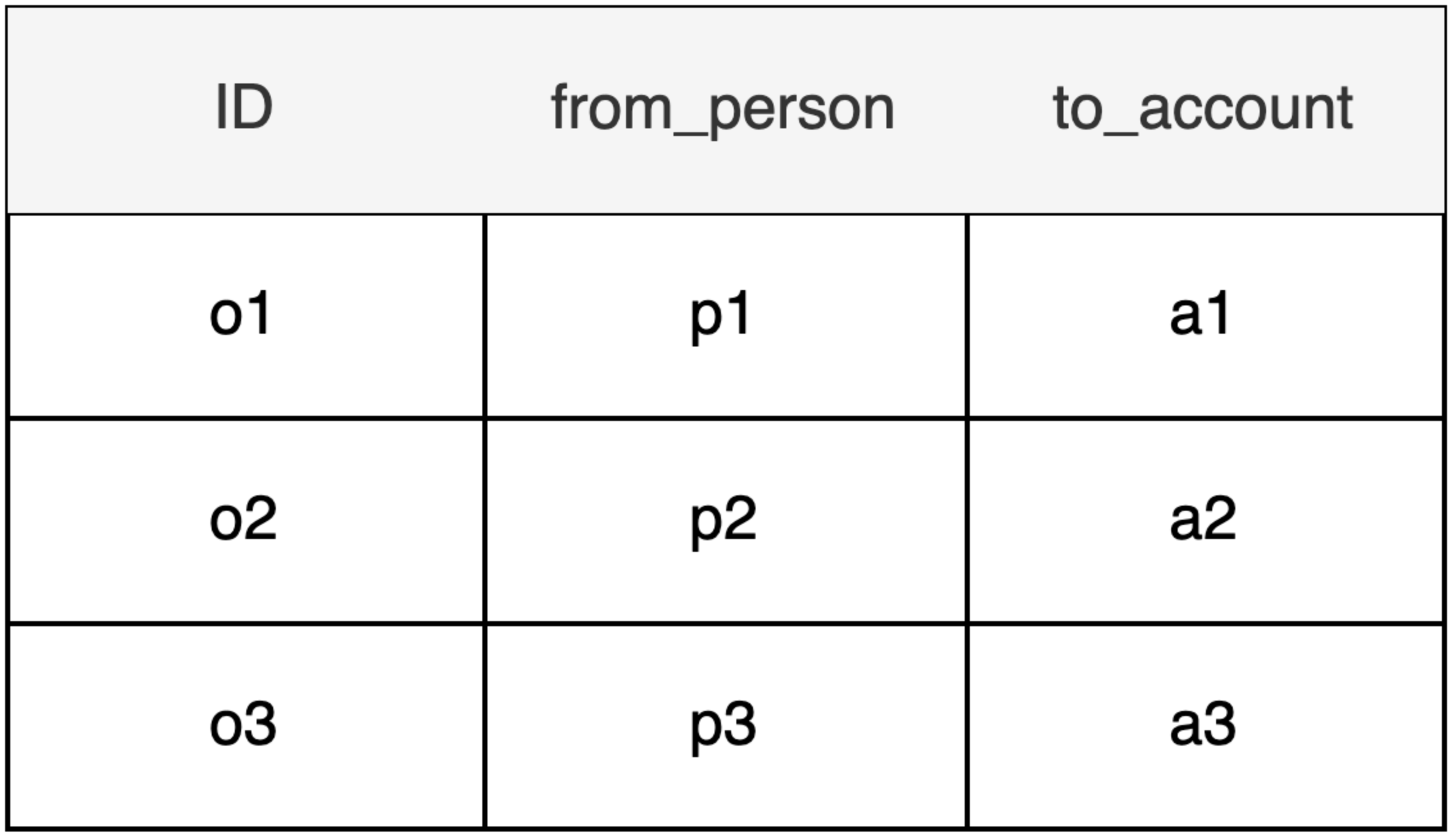}
        \caption{Own}
        \label{fig:own}
    \end{subfigure}
    \hfill
    \begin{subfigure}{0.25\textwidth}
        \centering
        \includegraphics[width=\linewidth]{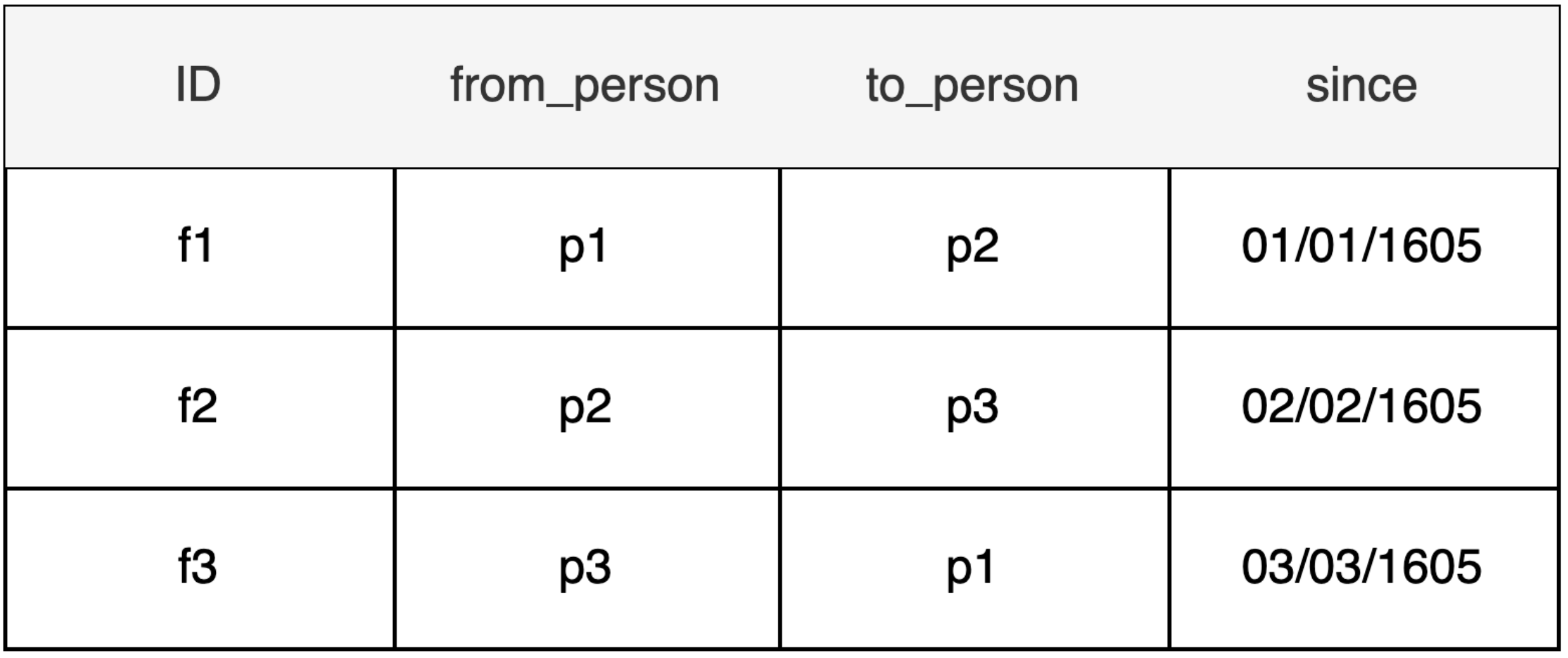}
        \caption{Friends}
        \label{fig:friends}
    \end{subfigure}
    \hfill
    \begin{subfigure}{0.28\textwidth}
        \centering
        \includegraphics[width=\linewidth]{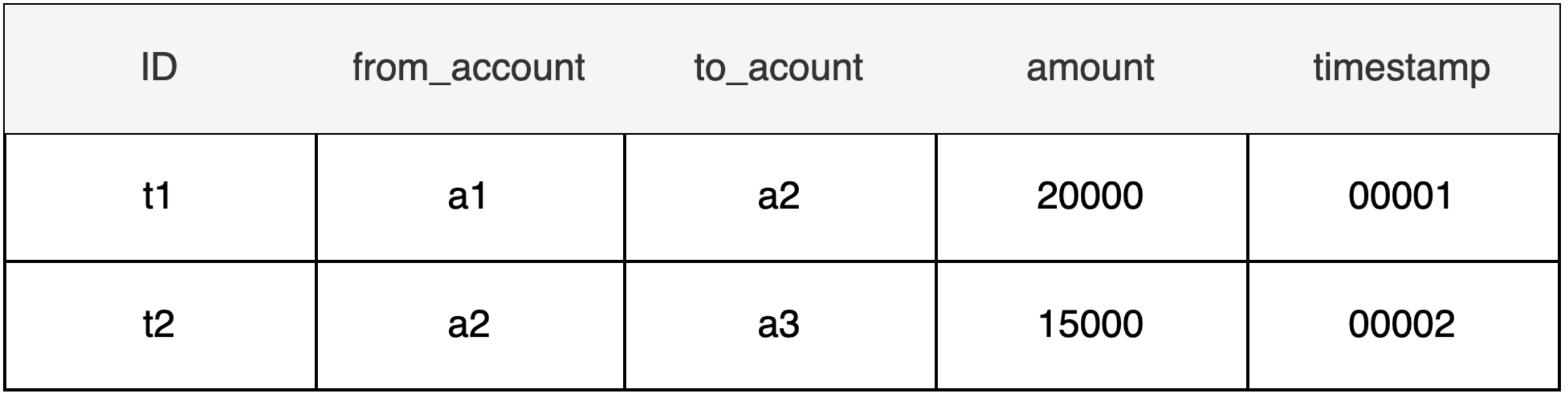}
        \caption{Transfer}
        \label{fig:transfer}
    \end{subfigure}
    
    \caption{Relational tables dataset with account, own, person, friends, and transfer. 
}
    \label{fig:combined}
\Description{}
\end{figure*}
\subsection{Experiments setup}
In our experiments, we implemented three distinct queries to evaluate latency differences between SQL and SQL/PGQ across different database interfaces, namely DuckDB, Google Cloud Spanner, and Neo4j. In this subsection, we elaborate on the specific implementation details and optimizations used in DuckPGQ, Google Cloud Spanner and Neo4j, highlighting their approaches to executing SQL/PGQ queries.

\subsubsection{DuckPGQ}
DuckPGQ is an extension module for DuckDB, designed specifically to integrate support for the SQL/PGQ syntax standardized in SQL:2023. DuckDB itself is a high-performance analytical database system optimized for single-node, embeddable analytics scenarios, featuring columnar storage, vectorized execution, and extensive extensibility via modular extensions.

DuckPGQ exploits DuckDB's extensibility mechanisms by introducing a parser extension capable of interpreting SQL/PGQ syntax, thus enabling native graph query functionality within DuckDB. The parsing stage translates SQL/PGQ-specific syntax such as the \texttt{MATCH} clause, inspired by Cypher, into logical relational algebra plans utilizing traditional relational operations and specialized scalar User Defined Functions (UDFs). These UDFs facilitate the creation of efficient Compressed Sparse Row (CSR) data structures dynamically, critical for performance-oriented graph traversal and pathfinding tasks.
The CSR data structures employed by DuckPGQ consist of two key components: a vertex array, indexing positions of outgoing edges, and an edge array, storing destination vertex positions. The motivation behind using dynamically generated CSR structures, rather than recursive SQL queries, stems from the need to minimize memory and computational overhead associated with path-finding operations. This structure is particularly advantageous in performing vectorized multi-source breadth-first search (MS-BFS) and shortest path calculations efficiently, leveraging DuckDB's built-in multi-core parallel execution capabilities and SIMD auto-vectorization.

DuckPGQ supports advanced graph queries including bounded and unbounded path searches, reachability queries, shortest path, and cheapest path computations. Particularly notable is DuckPGQ's implementation of cheapest path-finding via a SIMD-friendly variant of Multi-Source Bellman-Ford algorithm. Moreover, the extension enables flexible graph label management beyond standard SQL/PGQ definitions, allowing richer schema definitions with inheritance-like capabilities through discriminator columns.

Finally, the ongoing development roadmap for DuckPGQ includes further performance optimizations via worst-case optimal join (WCOJ) algorithms, improved parallel execution strategies, and expanded interoperability with graph-focused machine learning libraries (e.g., DGL and PyTorch Geometric), aiming to provide zero-copy data interchange between database operations and machine learning workflows.

\subsubsection{Google Cloud Spanner}
Google Cloud Spanner provides native support for graph data modeling through the Spanner Graph schema, based on the SQL/PGQ standard of SQL:2023. It models connected data as a network of nodes and edges, with nodes symbolizing the entities and edges representing the relationships between entities.

Google Cloud Spanner's explicit approach allows Cloud Spanner to efficiently manage and query relational graph data, ensuring optimized performance and seamless integration with Spanner’s scalable, distributed database capabilities.
Spanner Graph schema design involves defining nodes and edges explicitly through input tables. Node tables are standard Spanner tables where each row corresponds to a graph node, automatically inheriting the table name as the node label and table columns as node properties. 

Edges are modeled similarly through directed relationships, representing interactions such as ownership (\texttt{Owns}) or transactions (\texttt{Transfers}). By default, each directed edge explicitly encodes one-way relationships from source to destination nodes. Undirected edges, such as friendship, are represented using two directed edges with opposite directions.

\subsubsection{Neo4j}
Neo4j is a \textit{native graph database} purpose-built to store, query, and manage property graphs. According to the official Neo4j documentation~\cite{Neo4j_docs}, Neo4j organizes data as labeled nodes (representing entities) and directed, typed relationships (representing connections between entities), with both nodes and relationships holding properties in a key--value format.

\paragraph{Data Model and Query Language.}
Neo4j’s primary (and officially supported) query language is Cypher, a declarative language designed specifically for graph data.

Since there is no built-in way to run standard SQL in Neo4j, our experiments only include SQL/PGQ (or equivalently, Cypher-based) results. This is why, in the Results section ~\ref{sec:Results}, we do not provide entries for SQL on Neo4j.

Internally, Neo4j uses a cost-based query optimizer and specialized data structures for rapid graph traversal. The system maintains native adjacency lists, whereby each node references its outgoing (and incoming) relationships. This adjacency-focused storage format allows Neo4j to perform multi-hop path traversals efficiently without having to rely on a purely relational join model.
The typical workflow in Neo4j is to ingest or create data as graph structures: nodes labeled by domain concepts and relationships labeled by interaction types. These labeled nodes and relationships include properties to store relevant attributes.

When a user submits a Cypher query, Neo4j’s execution engine:
\begin{enumerate}
    \item Parses the query and transforms it into an abstract syntax tree (AST).
    \item Uses a cost-based planner to generate one or more potential execution plans, leveraging schema information, indexes (if present), and constraints to prune the search space.
    \item Applies optimization rules (e.g., pushing down predicates, inlining filters) to produce a final plan.
    \item Executes the plan using a runtime engine optimized for graph operations, such as expanding relationship chains, traversing neighbors, and filtering by labels or property values.
\end{enumerate}

Neo4j also incorporates caching strategies, storing frequently accessed graph structures and query plans in memory. When combined with its native transaction support (ACID-compliant), this architecture allows Neo4j to handle highly connected data patterns efficiently, outperforming many relational engines in scenarios involving multi-hop traversals. Because of its dedicated focus on graph operations, Neo4j remains one of the most popular platforms for graph use cases that involve complex relationship queries or real-time graph exploration.

\subsection{Experiments results}
\label{sec:experiments_results}

\begin{table}[ht]
\centering
\begin{tabular}{|l|c|c|c|c|c|c|c|c|}
\hline
\textbf{Language} & \makecell{\textbf{Dataset Size} \\ \textbf{(\# rows)}} & \textbf{Graph Creation (ms)} & \textbf{Q1 (ms)} & \textbf{Q2 (ms)} & \textbf{Q3 (ms)} & \textbf{Q4 (ms)} & \textbf{Q5 (ms)} & \textbf{Q6 (ms)} \\ \hline
\multirow{3}{*}{SQL} 
    & 50  & N/A    & 31.9  & 17.9   & 16     & 982   & 41.8    & 38.3  \\ \cline{2-9}
    & 100 & N/A    & 8.5   & 25.2  & 20.8   & 19.6  & 36.3m   & 36.9  \\ \cline{2-9}
    & 150 & N/A    & 12.4  & 19.2   & 19.9   & 371   & 12,420  & 70    \\ \hline
\multirow{3}{*}{SQL/PGQ}
    & 50  & 0.2  & 28.8  & 10.4   & 12.5   & 23.6  & 16.5    & 20    \\ \cline{2-9}
    & 100 & 0.4  & 25.6  & 11.7   & 10.6   & 61    & 43.3    & 18.9  \\ \cline{2-9}
    & 150 & 0.5  & 7.7   & 20.2   & 15.3   & 48.1  & 55.5   & 67.7 \\ \hline
\end{tabular}
\caption{DuckDB Performance Results: Queries 1–6 (Bounded and Unbounded)}
\label{tab:unified}
\end{table}

\begin{table}[ht]
\centering
\begin{tabular}{|l|l|l|l|l|l|l|}
\hline
\textbf{Language} & \makecell{\textbf{Dataset Size} \\ \textbf{(\# rows)}} & \textbf{Type} & \textbf{Graph Creation (ms)} & \textbf{Q1 (ms)} & \textbf{Q2 (ms)} & \textbf{Q3 (ms)} \\ \hline
{SQL} & 50 & Bounded & N/A & 15 & 29 & 44 \\
 & 100 & Bounded & N/A & 17 & 33 & 51 \\
 & 150 & Bounded & N/A & 18 & 29 & 39 \\
\hline
{SQL/PGQ} & 50 & Bounded & 4,700 & 30 & 177 & 1,020 \\
 & 100 & Bounded & 4,780 & 51 & 233 & 326 \\
 & 150 & Bounded & 4,870 & 37 & 179 & 866 \\
\hline
\end{tabular}
\caption{Google Spanner Performance Results: Bounded Queries (1-3)}
\end{table}

\begin{table}[ht]
\centering
\begin{tabular}{|l|c|c|c|c|c|c|c|c|}
\hline
\textbf{Language} & \makecell{\textbf{Dataset Size} \\ \textbf{(\# rows)}} & \textbf{Graph Creation (ms)} & \textbf{Q1 (ms)} & \textbf{Q2 (ms)} & \textbf{Q3 (ms)}  & \textbf{Q4 (ms)} & \textbf{Q5 (ms)} & \textbf{Q6 (ms)} \\ \hline
\multirow{3}{*}{Cypher}
    & 50  & 4,614 & 64  & 199 & 348 & 38  & 80  & 109 \\ \cline{2-9}
    & 100 & 4,668 & 113 & 197 & 405 & 53  & 113 & 109 \\ \cline{2-9}
    & 150 & 4,293 & 94  & 169 & 558 & 121 & 199 & 217 \\ \hline
\end{tabular}
\caption{Neo4j Performance Results (Cypher): Queries 1–6 (Bounded and Unbounded)}
\label{tab:neo4j-unified}
\end{table}

\begin{figure*}[ht]
\centering
\begin{subfigure}{0.3\textwidth}
    \includegraphics[width=\textwidth]{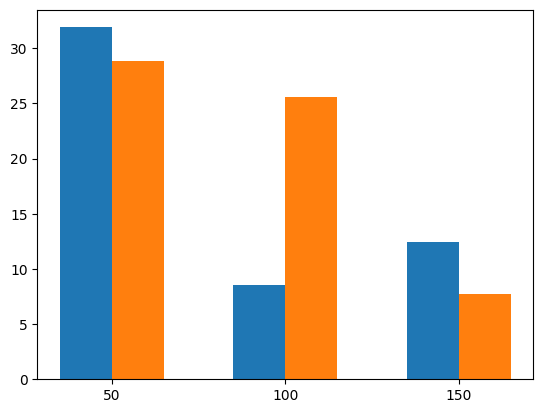}
    \caption{Q1}
    \label{fig:duckdb_query1_bounded_hist}
\end{subfigure}
\hfill
\begin{subfigure}{0.3\textwidth}
    \includegraphics[width=\textwidth]{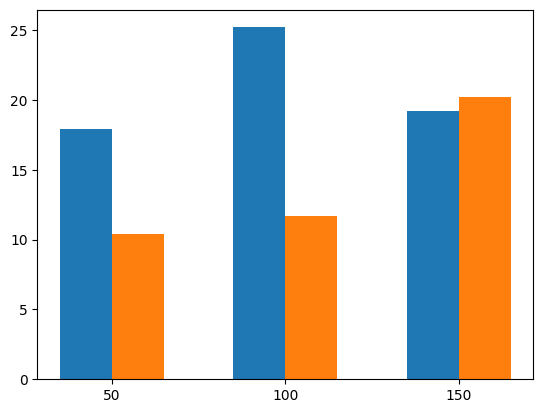}
    \caption{Q2}
    \label{fig:duckdb_query2_bounded_hist}
\end{subfigure}
\hfill
\begin{subfigure}{0.3\textwidth}
    \includegraphics[width=\textwidth]{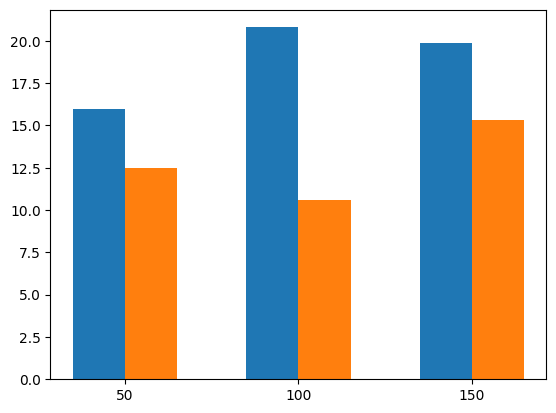}
    \Description{duckdb_query3_bounded_hist}
    \caption{Q3}
    \label{fig:duckdb_query3_bounded_hist}
\end{subfigure}
\begin{subfigure}{0.1\linewidth}
    \includegraphics[width=\linewidth]{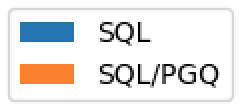}
\end{subfigure}
\caption{Comparison of execution time (ms) as a function of dataset size (\#rows) for SQL/PGQ queries for DuckDB (1-3 bounded queries)}
\label{fig:duckdb_bounded_hist_all}
\end{figure*}

\begin{figure*}[ht]
\centering
\begin{subfigure}{0.3\textwidth}
    \includegraphics[width=\textwidth]{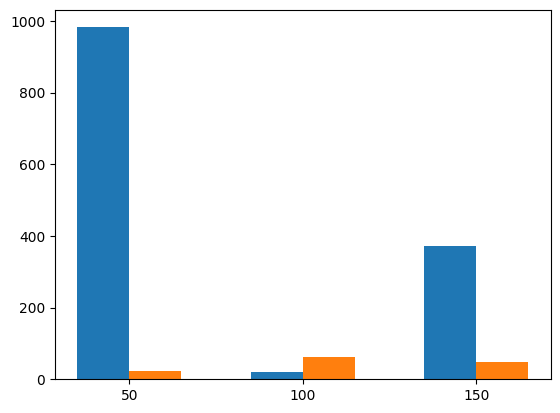}
    \caption{Q4}
    \label{fig:duckdb_query1_unbounded_hist}
\end{subfigure}
\hfill
\begin{subfigure}{0.3\textwidth}
    \includegraphics[width=\textwidth]{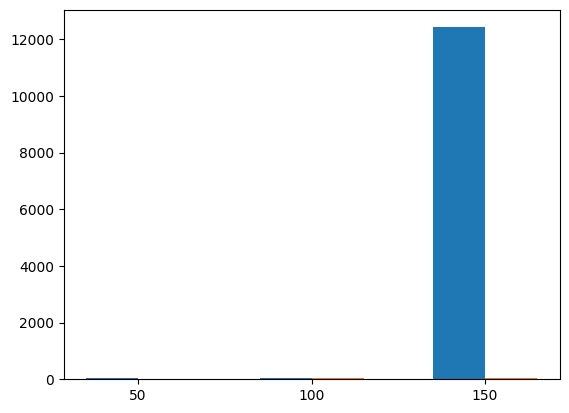}
    \caption{Q5}
    \label{fig:duckdb_query2_unbounded_hist}
\end{subfigure}
\hfill
\begin{subfigure}{0.3\textwidth}
    \includegraphics[width=\textwidth]{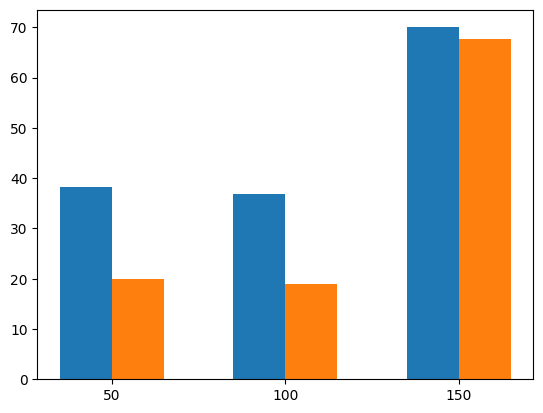}
    \Description{duckdb_query3_unbounded_hist}
    \caption{Q6}
    \label{fig:duckdb_query3_unbounded_hist}
\end{subfigure}
\begin{subfigure}{0.1\linewidth}
    \includegraphics[width=\linewidth]{plots/sql_pgq_hist_legend.png}
\end{subfigure}
\caption{Comparison of execution time (ms) as a function of dataset size (\#rows) for SQL/PGQ queries for DuckDB (4-6 unbounded queries)}
\label{fig:duckdb_unbounded_hist_all}
\end{figure*}

\begin{figure*}[ht]
    \centering
    \includegraphics[width=0.5\textwidth]{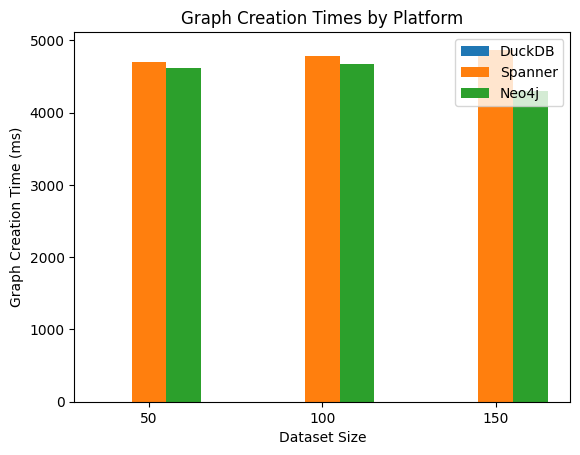}
    \Description{graph_creation_times}
    \caption{Comparison of graph creation times (ms) as a function of dataset size (\# rows)}
    \label{fig:graph_creation_times}
\end{figure*}

\begin{figure*}[ht]
\centering
\begin{subfigure}{0.3\textwidth}
    \includegraphics[width=\textwidth]{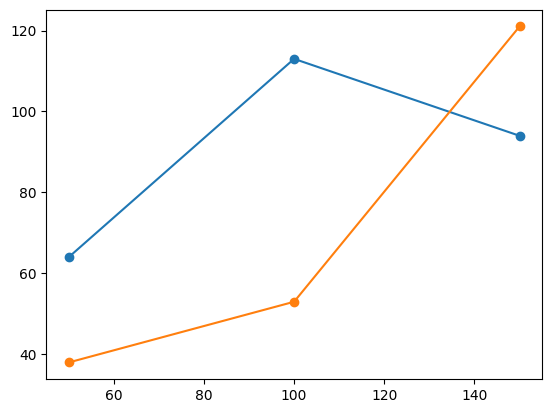}
    \Description{neo4j_query1}
    \caption{Q1 vs. Q4}
    \label{fig:noe4j_query1}
\end{subfigure}
\hspace{0.5em}
\begin{subfigure}{0.3\textwidth}
    \includegraphics[width=\textwidth]{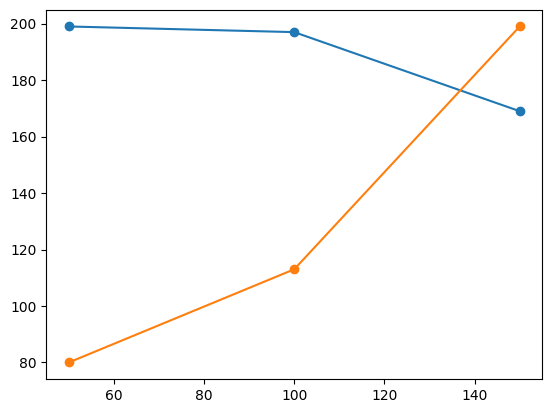}
    \caption{Q2 vs. Q5}
    \label{fig:noe4j_query2}
\end{subfigure}
\hspace{0.5em}
\begin{subfigure}{0.3\textwidth}
    \includegraphics[width=\textwidth]{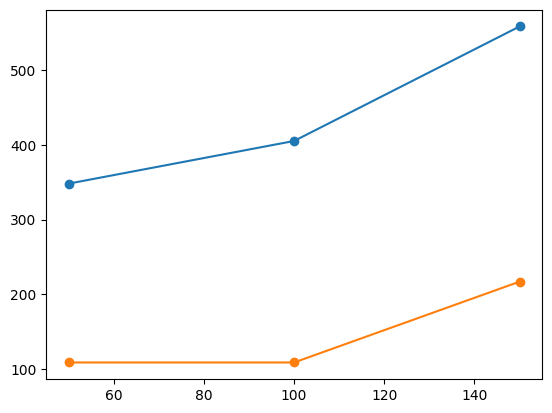}
    \caption{Q3 vs. Q6}
    \label{fig:noe4j_query3}
\end{subfigure}
\begin{subfigure}{0.1\linewidth}
    \includegraphics[width=\linewidth]{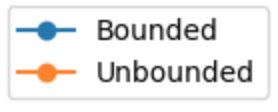}
\end{subfigure}
\caption{Comparison of execution time (ms) as a function of dataset size (\#rows) for Cypher queries for Neo4j (1-3 bounded queries)}
\label{fig:neo4j_queries}
\end{figure*}
\end{document}